\documentclass[twocolumn,showpacs,preprintnumbers,amsmath,amssymb]{revtex4}

\usepackage{graphicx}
\usepackage{amsmath,amsfonts}
\begin{document}

\author{David Kleinhans} \author{Rudolf Friedrich}
\affiliation{Westf{\"a}lische Wilhelms-Universit{\"a}t M{\"u}nster,
  D-48149 M{\"u}nster, Germany}

\author{Matthias W\"achter} \author{Joachim Peinke}
\affiliation{ForWind Center for Wind Energy Research,
  Carl-von-Ossietzky University of Oldenburg, D-26111 Oldenburg,
  Germany}

\title{Markov properties in presence of measurement noise}
\date{\today}


\begin{abstract}
  Recently, several powerful tools for the reconstruction of
  stochastic differential equations from measured data sets have been
  proposed [e.g. Siegert et al., Physics Letters A {\bf 243}, 275
  (1998); Hurn et al., Journal of Time Series Analysis {\bf 24}, 45
  (2003)]. Efficient application of the methods, however, generally
  requires Markov properties to be fulfilled. This constraint
  typically seems to be violated on small scales, which frequently is
  attributed to physical effects. On the other hand, measurement noise
  such as uncorrelated measurement and discretization errors has large
  impacts on the statistics of measurements on small scales.
  We demonstrate, that the presence of measurement noise, likewise,
  spoils Markov properties of an underlying Markov process. This
  fact is promising for the further development of techniques for the
  reconstruction of stochastic processes from measured data, since
  limitations at small scales might stem from artificial noise sources
  rather than from intrinsic properties of the dynamics of the
  underlying process. Measurement noise, however, can be controlled
  much better than the intrinsic dynamics of the underlying process.

\end{abstract}

\maketitle

\section{\label{sect:introduction}Introduction}
Physical systems often are described by means of dynamical systems
defined by differential equations of first order in time.  The
knowledge of a single point in phase space is sufficient for precise
prediction of the future evolution of the system. Starting from this
initial condition, the equations of motion can be integrated -- at
least numerically. Some systems are very sensitive to the initial
condition and therefore are associated with deterministic chaos.

For complex systems, a deterministic description often is not feasible
due to the huge amount of degrees of freedom and their frequently
unknown microscopic interactions. However, in many cases the
individual processes act on two different time scales. The dynamics of
the entire system then can be reduced to the dynamics of some
macroscopic order parameters, that enslave the highly fluctuating
microscopic degrees of freedom \cite{Haken:Synergetics}. In turn, the
set of order parameters, $\boldsymbol{x}$, obeys stochastic
differential equations (SDEs).  If the SDEs are of first order in
time, trajectories likewise can be generated from one single initial
state. The evolution then does not depend on properties of the
trajectory prior to the initial point and, therefore, exhibits only a
very restricted memory. Realisations of particular trajectories
sensitively depend on the fluctuating random forces, that are
involved.  However, considering an ensemble of realizations of the
stochastic process, the Markovian property becomes evident.

In recent years, the analysis of stochastic time series has made great
advances. Especially, the non-parametric reconstruction of the
governing stochastic differential equation by means of the direct
evaluation so drift and diffusion function has become a successful
tool for analyzing stochastic processes.  A method, that initially was
proposed by Siegert et al.~\cite{Siegert98}, in the meantime has been
applied to several problems in the field of finance
\cite{Friedrich00PRL}, life sciences
\cite{Kriso02,Sura,Kuusela04,Waechter04epj,Jafari03,Ivanova06,Prusseit07} and
turbulence \cite{Friedrich97}.  Moreover, algorithms for the efficient
application of maximum likelihood methods have been developed
\cite{Akaike81,Sahalia02}. A brief overview over the estimation power
of several methods can be found in \cite{Hurn03}. Quite recently, an
algorithm has been proposed, that combines the capabilities of the
latter methods \cite{Kleinhans05,Kleinhans07MLE}.  However, the
validity of Markov properties remains a crucial constraint for the
efficient application of all these procedures on stationary time
series data.

A close inspection of data sets generally indicates, that Markov
properties are violated at small time differences.  Typically,
physical arguments are accounted for this effect, based on the fact
that stochastic forces actually are correlated in time on small time
differences. The aim of the present note is to study the influence of
measurement noise on the Markov properties of measured data. We shall
show that measurement noise as well interferes with and spoils the
Markov properties.

The paper is organized as follows. In the next section, some methods
for verification of the Markov properties of measured data sets are
reconsidered. Section \ref{sect:theory} contains the basic arguments
concerning the influence of measurement noise on the transition
probability density functions. Consequences of the central equation
(\ref{eqn:markovnoise}) for the Markov properties will be made
explicit by means of three limiting cases, that are discussed at the
end of the section.  In section \ref{sect:data-analysis}, the general
results of the former section are exemplified by means of two
particular examples. In detail, the impact of discretization noise on
a purely deterministic system and the effects of uncorrelated
measurement noise on a stochastic process are investigated.  We
conclude with section \ref{sect:conclusion}, which summarises the main
results of our investigations and comprises the consequences for
standard tools for data analysis.

\section{\label{sect:markov}Verification of Markov properties}
Multivariate joint probability density functions (PDFs) are of great
importance for the analysis of measured time series
$\boldsymbol{x}(t)$. In principle, they contain all information on the
initial data set such as spatial and temporal evolution. The benefit
from a probabilistic approach on the basis of high dimensional joint
PDFs, however, generally is limited.

The analysis substantially can be simplified, if the data set under
consideration satisfies Markov properties.  This circumstance is
equivalent to the representation of all multivariate joint PDFs in
products of single-conditioned PDFs,
\begin{eqnarray}
  &P_n\left(\boldsymbol{x_n},\boldsymbol{x_{n-1}},\ldots,\boldsymbol{x_0}\right)=P_2\left(\boldsymbol{x_n}|\boldsymbol{x_{n-1}}\right)\nonumber\\&\times
  \ldots\times
  P_2\left(\boldsymbol{x_1}|\boldsymbol{x_{0}}\right) P_1\left(\boldsymbol{x_{0}}\right)\quad.
\end{eqnarray}
Here, $P_1\left(\boldsymbol{x_{i}}\right)$ is a shorthand notation for
the probability of being at time $t_i$ in a small interval at
$\boldsymbol{x_i}$ with $t_i<t_{i+1}\ \forall i$.  In general, the
latter transition PDFs furthermore explicitly depend on the times
$t_n,\ldots,t_0$.

Let us now assume the sample to be ergodic and stationary in a sense,
that ensemble averages can be carried out by means of time averages
and the PDFs do not depend on time explicitly \cite{Gardiner}.  Then,
the Chapman-Kolmogorov equation \cite{Risken,Gardiner}
\begin{equation}
  \label{eqn:chap-kol}
  P(\boldsymbol{x_{i}}|\boldsymbol{x_{i-2}})=\int d\boldsymbol{x_{i-1}} P(\boldsymbol{x_{i}}|\boldsymbol{x_{i-1}})P(\boldsymbol{x_{i-1}}|\boldsymbol{x_{i-2}})\quad,
\end{equation}
has to be fulfilled for any Markov process. This equation can be
evaluated numerically for measured data sets. Although the validation
of this equation is not sufficient for the validity of Markov
properties, it has turned out to be a very robust criterion.

\begin{figure*}
  \begin{center}
    \includegraphics*[width=7cm]{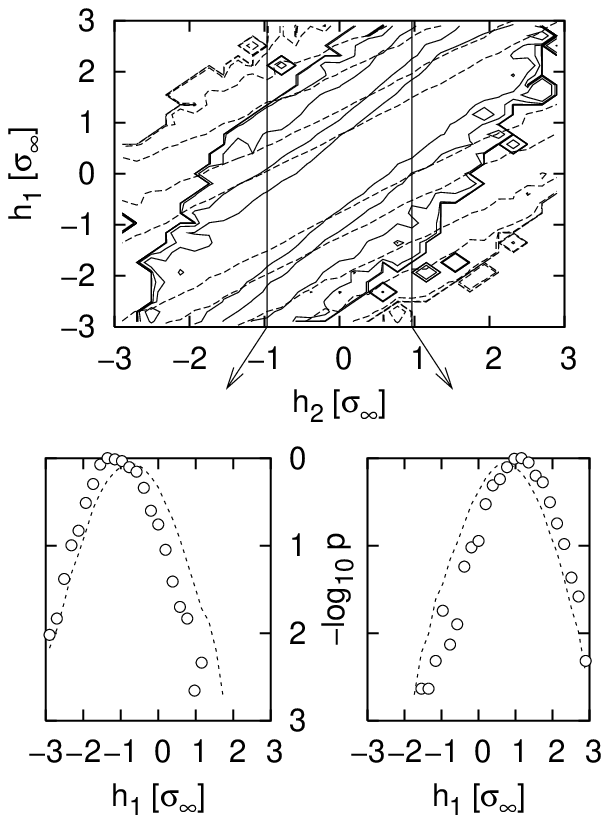}~\hspace*{1cm}~%
    \includegraphics*[width=7cm]{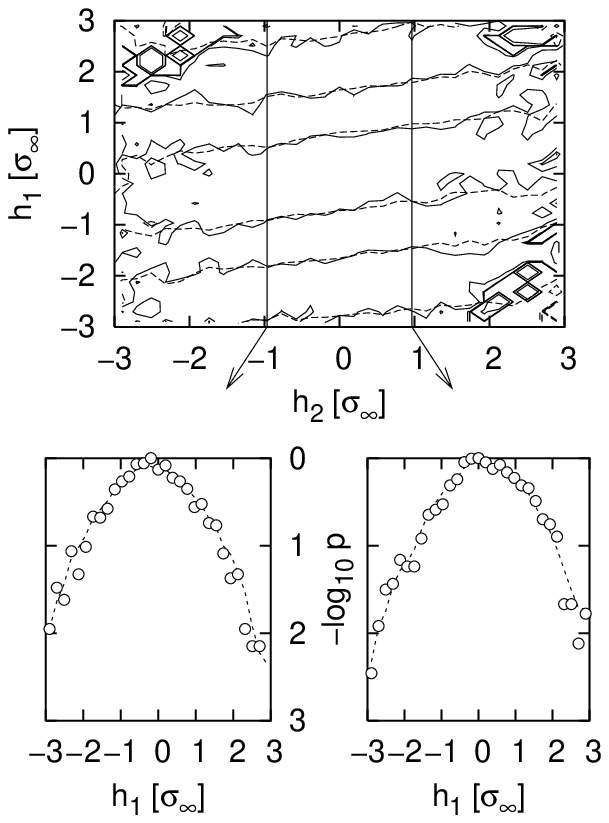}
  \end{center}
  \caption{\label{fig:markov_surf}Example for the analysis of Markov
    properties by means of graphical inspection of transition PDFs,
    prepared by W\"achter et al.~\cite{Waechter04epj}.  Test for
    Markov properties of Au film data for two different scale
    separations $\Delta r=14\ nm$ (lhs) and $35\ nm$ (rhs), where
    \mbox{$\Delta r=r_3-r_2=r_2-r_1$}.  In both cases $r_2=169\ nm$.
    In each case a contour plot of conditional probabilities
    $P(h_1,r_1 | h_2,r_2)$ (dashed lines) and $P(h_1,r_1 |
    h_2,r_2;h_3\!\!=\!\!0,r_3 )$ (solid lines) is shown in the top
    panel.  Contour levels differ by a factor of 10, with an
    additional level at $p=0.3$.  Below the top panels in each case,
    two one-dimensional cuts at $h_2 \approx \pm \sigma_\infty$ are
    shown with $P(h_1,r_1 | h_2,r_2)$ as dashed lines and $P(h_1,r_1 |
    h_2,r_2;h_3\!\!=\!\!0,r_3 )$ as circles. From the deviations of
    the PDFs for $\Delta r=14\ nm$ (lhs) it becomes evident, that
    Markov properties are not fulfilled in this case. They might,
    however, be valid for $\Delta r=35\ nm$ (rhs).}
\end{figure*}

Moreover, a direct comparison of the conditional probability
distributions $P({\bf x}_2|{\bf x}_1,{\bf x}_0)$ and $P({\bf x}_2|{\bf
  x}_1)$ has been used for validation of Markov properties. For
Markovian data, these functions should coincidence for arbitrary
values of ${\bf x}_0$. An example for the application of this
procedure by means of graphical inspection of the PDFs is depicted in
figure \ref{fig:markov_surf}, that has been prepared by W\"achter et
al.~in connection with the study of the statistical properties of
hight profiles of gold surfaces \cite{Waechter04epj}.  Here, W\"achter
et al.~investigated Markov properties of the transition PDFs for
nested heigth increments in different scales. In the present case,
Markov properties might be fulfilled for scales separated by $\Delta
r=35\ nm$, whereas they evidently are violated at separation lenghts
of $\Delta r=14\ nm$, as can be seen from inspection of figure
\ref{fig:markov_surf}.  It is evident, that the proper interpretation
of the plot with respect to the Markov properties has to be quantified
by introducing a certain measure for the distance of the two
probability distributions.  To this end the Wilcoxon test
\cite{Wilcoxon45,Mann47} can be applied in order to compare PDFs, that
originate from samples of different size, and only makes few demands
on the properties of the individual PDFs.  The numerical
implementation is straightforward, results for the present example
e.g.~are depicted in \cite{Waechter04epj}. For a detailed description
of the Wilcoxon test we refer to the appendix of \cite{Renner01}.

If the direct estimation \cite{Siegert98} of drift and diffusion
functions from measured data sets is intended and the underlying
process, therefore, is assumed to obey Langevin equations, an
alternative method can be applied for inspection of Markov properties.
Once the estimation procedure has been performed and an estimate for
drift and diffusion functions is available, the character of the
dynamical noise can be determined from the sample.  The presence of
noise without any temporal and spatial correlations is a sufficient
indication for compliance of the measured data set with Markov
properties. This procedure e.g.~is outlined and applied in
\cite{Marcq01}. It is certainly the most direct way to investigate
Markov properties.

\section{\label{sect:theory}Impact of measurement noise on Markov
  properties}
An ensemble of Markov processes $x(t)$ is considered, that now is
distorted by measurement noise $\xi(t)$. For simplicity, the details
are carried out for a one-dimensional process. Only three consecutive
points $x_0$, $x_1$ and $x_2$ with $x_i:=x(t_i)$ and $t_i:=t_0+i\tau$
are investigated for this purpose.  Since the statistics is assumed to
be stationary, this is sufficient for the current considerations.
Henceforth, $P_x(x_{i+1}|x_i)$ is a shorthand notation for the
transition PDF of the variable $x$ in the time increment $\tau$.

Let us now assume, that the true process is hidden to the data
analyst: Instead of the variable $x(t)$, a perturbed variable $y(t)$
is measured, that emerges from the initial process by means of the
relation
\begin{equation}
  y(t)=x(t)+\xi(x(t),t)\quad.
\end{equation}
Thereby, $\xi(x(t),t)$ is a stochastic variable, that incorporates
systematic and non-systematic measurement errors. We further assume,
that the deterministic contributions to the measurement error can be
identified and the noise $\xi$ can be specified by
\begin{equation}
  \xi(x(t),t)=\xi_s(x(t)+\xi_{ns}(t))\quad.
\end{equation} 
Here, $\xi_{ns}$ incorporates non-systematic noise sources. For
reasons of simplicity, we assume these errors to be independent of one
another for consecutive measurements,
\begin{equation}
  \left\langle\xi_{ns}(t+\tau)\xi_{ns}(t)\right\rangle\sim\delta(\tau)\quad.
\end{equation}
On the other hand, $\xi_s$ characterises deterministic, systematic
measurements errors, that have no explicit dependence on $t$. We would
like to emphasize, that discretization errors fall into this broad
class, that are an intrinsic feature of any digital measurement
procedure. While the former noise is uncorrelated, this assumption
generally is violated for the latter noise source due to correlations
in the variable $x$ itself.
 
The probability for the measurement of $y_i$ now solely depends on the
entangled variable $x_i$ and can be specified by means of the
conditional probability $P_{\xi}(y_i|x_i)$.  Hence, the conditional
probability $P_y(y_2|y_1,y_0)$ for the process $y(t)$ can be
calculated by means of its definition through joint probabilities.
Application of the Markov properties of the underlying process $x(t)$
finally yields
\begin{widetext}
  \begin{equation}
    P_y(y_2|y_1,y_0)=\frac{P_y(y_2,y_1,y_0)}{P_y(y_1,y_0)}=\frac{\int dx_2\int dx_1\int dx_0\ P_\xi(y_2|x_2)P_\xi(y_1|x_1)P_\xi(y_0|x_0)P_x(x_2|x_1)P_x(x_1|x_0)P_x(x_0)}{\int dx_1\int dx_0\ P_\xi(y_1|x_1)P_\xi(y_0|x_0)P_x(x_1|x_0)P_x(x_0)}\label{eqn:markovnoise}
  \end{equation}
\end{widetext}
In general, this expression deviates from the single conditioned PDF
$P_y(y_2|y_1)$. Therefore, noisy measurements on perfect Markov
processes in general lose their Markov property due to the inexact
measurement procedure.

Referring to section \ref{sect:introduction} this means, that a single
point from a noisy measurement on a Markov process, $y(t_n)$, not in
any case is sufficient for a proper prediction of the future dynamics
of the measured data. This makes sense, since the intrinsic state of
the system, $x(t_n)$, hardly can be estimated from just one single
measurement due to the measurement uncertainty. Rather, the
consideration of a couple of noisy measurements, $y(t_0), \ldots,
y(t_n)$, can enhance the accuracy of the predicted probability of
$y(t_{n+1})$.

At least for three simple cases, expression (\ref{eqn:markovnoise})
can be investigated analytically.

First, Markov properties are retrieved for the trivial case
$P_\xi(y|x)=\delta(y-x)$, where actually no measurement noise is
present.

Second, (\ref{eqn:markovnoise}) can be evaluated for
$P_x(x_{i+1}|x_i)=P_x(x_{i+1})$. In this case, the entangled process
itself does not show any correlations.  Frequently, this approximately
is true for large time increments between individual measurements.
If so, the integrals disentangle and the noisy measurements themselves
turn out to be independent of one another,
$P_y(y_2|y_1,y_0)=P_y(y_2)$. Thus, the measured variable $y$ satisfies
Markov properties.

Third, noisy measurements can be considered, that sample the process
much faster than the intrinsic dynamics of the entangled variable,
$x$. Therefore, $P_x(x_2|x_1)=\delta(x_2-x_1)$ is a reasonable
approximation of the transition PDF on consecutive measurements.
Moreover, only purely non-systematic, Gaussian measurement noise with
variance $\sigma^2$ is taken into account. In this case, evaluation of
expression (\ref{eqn:markovnoise}) yields
\begin{eqnarray}
  & P_y(y_2|y_1,y_0)=\sqrt\frac{1}{3\pi\sigma^2}\exp\left[-\frac{\left(2y_2-y_1-y_0\right)^2}{12\sigma^2}\right]\\\nonumber&\times\frac{\int dx_0\ \sqrt\frac{1}{2\pi\sigma^2/3}\exp\left[-\frac{\left(x_0-\frac{1}{3}\left(y_2+y_1+y_0\right)\right)^2}{2\sigma^2/3}\right]P_x(x_0)}{\int dx_0\ \sqrt\frac{1}{2\pi\sigma^2/2}\exp\left[-\frac{\left(x_0-\frac{1}{2}\left(y_1+y_0\right)\right)^2}{2\sigma^2/2}\right]P_x(x_0)}\quad.
\end{eqnarray}
In the latter factor, two different convolutions occur in numerator
and denominator: The stationary PDF $P_x(x_0)$ is convoluted with
Gaussian PDFs with different standard deviations, centred at the
average value of $y_2,y_1,y_0$ and $y_1,y_0$, respectively. Therefore,
this expression generally depends on the value $y_0$ and conflicts
with Markov properties of $y(t)$. We would like to emphasize, that the
approximation of a persistent entangled process is feasible for fast
but noisy measurements on rather slow processes. The current case
reveals the loss of Markov properties on the very small time scales
for these kind of measurements, that does not stem from its intrinsic
dynamics but, purely, from uncertainties during the measurement
process.

\section{\label{sect:data-analysis} Examples}
Let us now elucidate the findings of the latter section by means of two
examples. First, the influence of discretization noise on the
properties of a simple deterministic process is investigated. By
construction, the violation of the Chapman-Kolmogorov equation can be
demonstrated.

Second, the influence of Gaussian measurement noise on the Markov
properties of a stochastic process at relatively high time lag is
considered. The effect of measurement noise becomes obvious from the
inspection of conditional PDFs obtained by numerical integration
of the Chapman-Kolmogorov equation (\ref{eqn:chap-kol}).

\subsection{Influence of discretization noise on a deterministic
  process}
We consider the elementary process
\begin{equation}
  \label{eqn:zerfall}
  x(t_0+\tau)=x(t_0)\exp\left[-\gamma\tau\right]\quad.
\end{equation}
It is the general solution of the ordinary differential equation
$\dot{x}=-\gamma x$. Since the dynamics are of first order in time,
the one dimensional process $x(t)$ can be specified by one initial
condition and, therefore, is Markovian. Due to the deterministic
character, the conditional transition PDF for the variable $x$ in the
time interval $\tau$ complies with
\begin{equation}
  \label{eqn:zerfall_x_cpdf}
  P_x(x_1|x_0,\tau)=\delta\left(x_1-x_0 e^{-\gamma\tau}\right)\quad.
\end{equation}
The process apparently is not stationary, since no forcing is present.
The conditional transition PDFs, however, do not depend on time
explicitly.  We now consider the statistics of an ensemble of
measurements, whose initial positions $x(t_0)$ of the individual
processes are distributed according to $P_x(x)$.

\begin{figure}
  \includegraphics*[width=8cm]{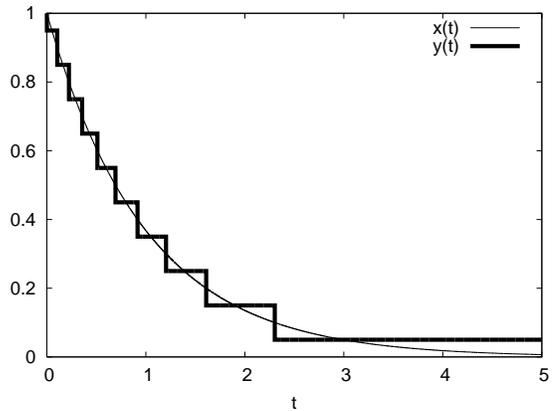}
  \caption{\label{fig:zerfall_diskret} Example of a process $x(t)$
    according to eqn.~(\ref{eqn:zerfall}), that is affected by strong
    discretization noise. Here, the faint line specifies the original
    process $x(t)$, whereas the bold line depicts the evolution
    $y(t)$, that eventually is obtained from the measurement due to
    discretization errors.}
\end{figure}

We assume that the exact intrinsic variable $x$ is entangled due to
discretization errors, that occur during an imaginary measurement
procedure.  Therefore, the exact, continuous variables $x$ are mapped
to a finite set of discrete variables
$\Omega=\{\omega_0,\ldots,\omega_n\}$ according to the rule
\begin{equation}
  \label{eqn:sample_map} x\rightarrow y=\omega_i\quad\mbox{such that} \quad \omega_i^-\le x< \omega_i^+\quad.
\end{equation} 
Here, the intervals $[\omega_i^-,\omega_i^+]$ and
$[\omega_{i+1}^-,\omega_{i+1}^+]$ associated with the variables
$\omega_i$ and $\omega_{i+1}$ are connected to one another by the
requirements $\omega_i^+=\omega_{i+1}^-$ and $\omega_i^-<\omega_i^+$.
Moreover it is implied, that any measured value $x$ can be mapped by
means of (\ref{eqn:sample_map}). The interval
$[\omega_0^-,\omega_n^+]$, thus, covers all values $x(t)$ that are
realised by any process under consideration at any time $t$. The
discretization noise can be specified in compliance with the notation
of the latter section by the conditional PDF
\begin{equation}
  \label{eqn:zerfall_xi_cpdf}
  P_\xi(y|x)=\left\{\begin{array}{lcl}1&\mbox{if}&x\in[y^-,< y^+[\\0&\mbox{if}&x\notin[y^-,< y^+[\end{array}\right.\quad.
\end{equation}
The effect of discretization noise on the initial variable $x$ is
illustrated in figure \ref{fig:zerfall_diskret}. As $y$ only assumes
discrete values $\omega_0,\ldots,\omega_n$, the normalisation of the
latter PDF for any $x$ is guaranteed by the equation
\begin{equation}
  \sum\limits_{y\in\Omega} P_\xi(y_i|x)=1\quad.
\end{equation}

We now would like to demonstrate the loss of Markov properties due to
the discretization of the signal.  In principle,
eqn.~(\ref{eqn:markovnoise}) directly could be evaluated numerically
for the ensemble under consideration. However, in this case the
invalidity of the Chapman-Kolmogorov equation (\ref{eqn:chap-kol})
nicely can be utilized for this purpose.

Analogous to eqn.~(\ref{eqn:markovnoise}), the transition PDF
conditioned on a single point can be specified,
\begin{eqnarray}
  \label{eqn:zerfall-transpdf1}
  &&P_y(y_1|y_0,\tau)\\\nonumber&&=\frac{\int dx_1\int dx_0\
    P_\xi(y_1|x_1)P_\xi(y_0|x_0)P_x(x_1|x_0,\tau)P_x(x_0)}{\int dx_0\ P_\xi(y_0|x_0,\tau)P_x(x_0)}\quad.
\end{eqnarray}
Application of the particular transition PDFs
(\ref{eqn:zerfall_x_cpdf}) and (\ref{eqn:zerfall_xi_cpdf}) yields
\begin{equation}
  P_y(y_1|y_0,\tau)=\frac{\int
_{\max\left(y_0^-,e^{\gamma\tau}y_1^-\right)}^{\min\left(y_0^+,e^{\gamma\tau}y_1^+\right)}dx_0\
    P_x(x_0)}{\int
_{y_0^-}^{y_0^+}dx_0\
    P_x(x_0)}\quad.
\end{equation}
If the process $y(t)$ would obey Markov properties, the discrete
version of the Chapman-Kolmogorov equation,
\begin{equation}
  \label{eqn:zerfall_chap-kol}
  P_y(y_2|y_0,2\tau)=\sum\limits_{y_1\in\Omega}P_y(y_2|y_1,\tau)P_y(y_1|y_0,\tau)\quad,
\end{equation}
would have to be fulfilled for any choice of $y_2$, $y_0$ and $\tau$.
For $y_2=y_0=y$ with $y^->0$ and
$\tau=\log\left(y^+/y^-\right)/(2\gamma)$, the invalidity of this
equation is evident, if $P_x(x)>0$ for $x\in\left[y^-,y^+\right]$: The
left hand side of eqn.~(\ref{eqn:zerfall_chap-kol}) vanishes, whereas
the sum on the right hand side involves the summand
\begin{equation}
  \left[ \frac{\int
_{e^{\log\left(y^+/y^-\right)/2}y^-}^{y^+}
      dx_0\
      P_x(x_0)}{\int
_{y^-}^{y^+}dx_0\
    P_x(x_0)}\right]^2>0\quad.
\end{equation} 
As the other summands are non-negative, the Chapman-Kolmogorov
equation is violated for the process under consideration.
Consequently, the distorted process $y(t)$ does not comply with Markov
properties any more.

\subsection{Influence of measurement noise on a stochastic process}
\begin{figure}
  \includegraphics*[width=8cm]{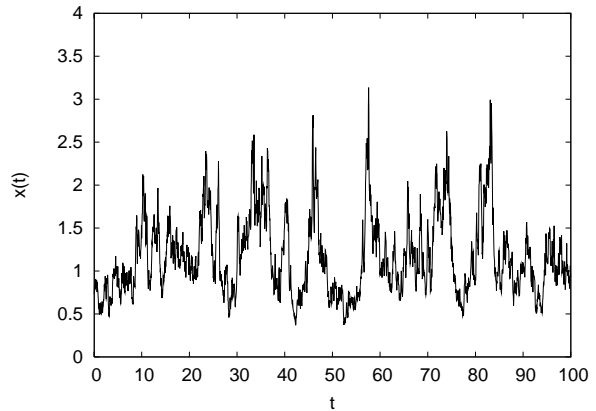}
  \caption{\label{fig:lognormal_sample} Detail of a sample path of the
    stochastic process (\ref{eqn:lognormal_drift_diff}) for the
    parameters $(\gamma,D)=(0.75,0.1)$. The occurrence of distinct
    peaks is characteristic for multiplicative stochastic processes.}
\end{figure}

\begin{figure*}
  \includegraphics*[width=15cm]{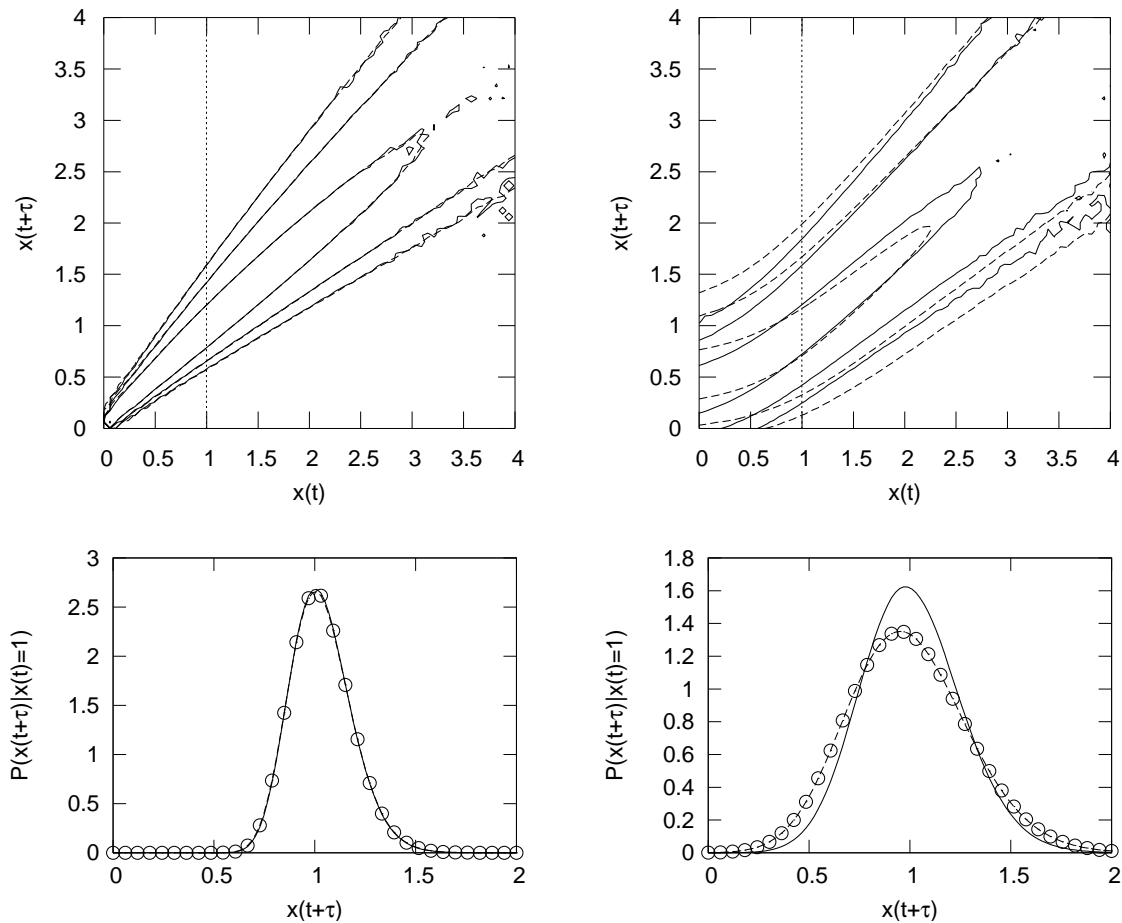}
  \caption{\label{fig:lognormal_cpdfs} Test for Markov properties for
    simulated samples A without measurement noise (lhs) and B with
    artificial Gaussian measurement noise with variance $2.25\cdot
    10^{-2}$ (rhs), respectively. In the upper panels, the conditional
    transition PDFs for $\tau=0.1$ (solid contour lines) are compared
    with the ones obtained for the same time increment by numerical
    integration of the Chapman-Kolmogorov equation
    (\ref{eqn:chap-kol}) for transition PDFs for increment $\tau/2$
    (dashed contour lines). Contour lines are placed at the levels
    $10$, $1$, $0.1$ and $0.01$. In the lower panels, a cross section
    of the transition PDF at $x(t)=1$ is depicted. For reasons of
    clearness, circles have been added to the dashed lines
    corresponding to the data set B. Perfect coincidence of the PDFs is
    observed for A, whereas in case of B systematic
    deviations become evident. Consequently, Markov properties are
    spoiled by the artificial measurement noise of sample B.}
\end{figure*}

The influence of Gaussian measurement noise on an one dimensional
stochastic process with drift and diffusion functions
\begin{subequations}
  \label{eqn:lognormal_drift_diff}
  \begin{eqnarray}
    D^{(1)}(x)&=&x\left(D-\gamma\log\left(\frac{x}{x_0}\right)\right)\\
    D^{(2)}(x)&=&D x^2
  \end{eqnarray} 
\end{subequations}
is investigated. For further details on stochastic processes we refer
to \cite{Gardiner, Risken}. This process already has been discussed in
\cite{Kleinhans07MLE} within the scope of an analytical example.
Thereby, the following procedure for the exact simulation of a
discrete sample of this process by means of the underlying
Ornstein-Uhlenbeck process $s(t)$ has been motivated,
\begin{subequations}
  \begin{eqnarray}
    x_i&=&\exp\left[s_i\right]\label{eqn:lognorm-beisp-simx}\\
    \label{eqn:lognorm-beisp-sims}
    s_{i+1}&=&e^{-\gamma \Delta
      t}s_i+\sqrt{\frac{D}{\gamma}(1-e^{-2\gamma \tau})}\Gamma_i\quad.
  \end{eqnarray} 
\end{subequations}
Here, equation (\ref{eqn:lognorm-beisp-sims}) is the rule for the
discrete simulation of an underlying Ornstein-Uhlenbeck process $s$,
where $\Gamma_i$ are normally distributed independent random variables
with variance 1. It is deduced from the transition PDFs for the
Ornstein-Uhlenbeck process, that exactly can be specified even for
finite time lag $\tau$ \cite{Risken}. In this vein, discretization
errors stemming from the standard schemes for the numerical
integration of SDEs \cite{Kloeden} are avoided. The starting value
$s_0$ should be drawn from a Gaussian distribution with variance
$D/\gamma$, which is the stationary distribution of the process $s$.
The desired process $x$ is obtained from the process $s$ by means of
the nonlinear transform (\ref{eqn:lognorm-beisp-simx}). A sample
process for parameter set $(\gamma,D)=(0.75,0.1)$ is depicted in
figure \ref{fig:lognormal_sample}.

For the current example, time series A consisting of $50\cdot 10^6$
sample points with time increment $\tau=0.05$ was simulated. A second
series, B, was generated from series A by addition of independent,
identically distributed Gaussian random variables with variance
$2.25\cdot 10^{-2}$, that model noise stemming from non-systematic
measurement errors. Both series A and B have been subjected to the
same analyzing procedure: Conditional PDFs have been calculated from
the for time increment $\tau=0.1$. On the other hand these conditional
PDFs have been calculated from conditional PDFs for the time lag
$\tau=0.05$ by means of numerical integration of the
Chapman-Kolmogorov equation, (\ref{eqn:chap-kol}).  The results are
exhibited in figure \ref{fig:lognormal_cpdfs}. In theory, these PDFs
should coincidence with the former ones for Markovian processes.
However, distinctive systematic deviations show up in presence of
measurement noise, as can be seen from the analysis of data set B in
figure \ref{fig:lognormal_cpdfs}. Hence, the artificial measurement
noise of time series B interferes with the Markov properties of the
underlying time series A.

The sets A and B correspond to the first and third limiting case of
equation (\ref{eqn:markovnoise}), respectively, that were discussed at
the end of section \ref{sect:theory}. The second case also can be
investigated by means of the current example with an increased time
lag $\tau$, such that $\exp(-\gamma\tau)\ll 1$. Then, Markov
properties are reobtained even in case of strong measurement noise.

\section{\label{sect:conclusion}Conclusion}
The influence of different noise sources on the structure of
multivariate joint probability distribution functions has been
investigated. In particular, the effects of noise on the sensitivity
of transition probability density functions to an additional, second
condition has been analysed. It turned out, that noise generally has
impacts on these transition probability density functions and
seriously interferes with Markov properties, even if they are
fulfilled for the original, uncorrupted process. This fact is, in our
opinion, counter-intuitive.

The analysis of samples, that are affected by measurement noise,
already for a long time is routine in applied sciences and industrial
applications. Typically, Kalman filtering is applied fur this purpose
\cite{Kalman60}. For a recent review on this and other iterative
techniques we refer to \cite{Voss04}. Recently, Siefert et
al.~\cite{Siefert03} addressed this problem from a dynamical systems'
point of view. The intention was to extend the efficient
non-parametric estimation procedure proposed by Siegert and
al.~\cite{Siegert98} to data suffering from measurement noise.  In
this context it could be shown, that intrinsic dynamical and external
measurement noise in principle can be separated from one another, if
the sampling frequency is sufficiently high whereas the amplitude of
the measurement noise is weak. Following, B\"ottcher et al.~succeeded
in the efficient reconstruction of simple processes even in presence
of strong measurement noise \cite{Boettcher06}.  Although the latter
work is based on eqn.~(\ref{eqn:zerfall-transpdf1}), the general
problem of the vanishing Markov properties in presence of measurement
noise could not be identified. This new point of view, however,
involves a broad class of tools that are available for data analysis,
since most tools rely on a finite embedding of the data.

The new insights have consequences for future analysis of time series:
The influence of measurement noise should be discussed for any
individual method, that is applied for the analysis of time series.
Explicitly, also effects stemming from discretization errors should be
considered here. Eventually, methods might be applicable even to data
sets, that until now could not be processed due to invalidity of
Markov properties.  This feature, however, might stem from artificial
noise rather than from intrinsic properties of the dynamics of the
underlying process.


\end{document}